\documentclass[aps,prl,reprint,superscriptaddress]{revtex4-2}

\usepackage{bm}
\usepackage{graphicx}
\usepackage{color}
\usepackage{amsmath}
\usepackage{hyperref}
\usepackage{epstopdf}
\usepackage{upgreek}
\usepackage{mathptmx, textcomp}
\usepackage[latin1]{inputenc}
\usepackage[T1]{fontenc}
\usepackage{array, multirow}
\usepackage[table]{xcolor}
\usepackage{booktabs}
\usepackage{longtable}
\usepackage{comment}

\begin{document}

\title{Precision probing of ionic-core transitions in alkaline-earth Rydberg atoms}

\author{Mitsuki Odahara}
\affiliation{Graduate School of Engineering Science, The University of Osaka, 1-3 Machikaneyama, Toyonaka, Osaka 560-8531, Japan}

\author{Shinsuke Haze}\email{shinsuke.haze.qiqb@osaka-u.ac.jp}
\affiliation{Center for Quantum Information and Quantum Biology, The University of Osaka, 1-2 Machikaneyama, Toyonaka, Osaka 560-0043, Japan}

\date{\today}

\begin{abstract}
We report precision spectroscopy of ionic-core transitions in alkaline-earth Rydberg atoms.
We demonstrate high-resolution measurements of isotope shifts and hyperfine splitting of dipole transitions in ionic cores which have not been explored so far.
A key element of this work is the reduction of the linewidth by more than two orders of magnitude enabled by dynamical control of Rydberg electron's orbit which significantly enhances the spectral resolution.
Furthermore, to unambiguously identify the frequency shift, we directly compare core ion's spectrum with a signal from a single trapped ion serving as an ultimate frequency reference.
This work provides an important foundation for quantum control of inner-core transitions, which offer an useful tool in manipulating Rydberg atom as well as a sensitive probe for electron-core interactions in atomic and molecular systems.

\end{abstract}

\maketitle


Precision spectroscopy and quantum control of atomic and molecular degrees of freedom at an unprecedented level lies at the heart of fundamental physics and quantum science applications.
Such ultra-precise spectroscopy can, for example, be used for testing general relativity \cite{Takamoto} and searches for new bosons \cite{Counts} as well as atomic clocks application \cite{Ludlow}.
In these applications, narrow-line optical transitions of atoms and molecules serve as ideal targets for precision spectroscopy.
For instance, optical lattice clocks \cite{Aeppli} and single-ion optical clocks \cite{Marshall} have demonstrated remarkable progress, establishing a new frontier in frequency metrology.

Ionic cores embedded in Rydberg atoms, e.g., alkaline-earth metals, have recently emerged as intriguing and promising platforms for quantum control and precision spectroscopy.
In particular, inner-core optical transitions are found to be highly sensitive to perturbations from the outer electron, making it a sensitive probe of electron-core interactions in atomic systems \cite{Camus, Warntjes, Marin-Bujedo}.
In addition, ionic cores in Rydberg atoms have recently attracted attention as they can be used for trapping, detecting \cite{Lochead, Madjarov,McQuillen}, and controlling \cite{Burgers,Pham} of Rydberg atoms.
More recently, intriguing proposals, including direct laser cooling of Rydberg atoms via ionic-core transitions \cite{Bouillon}, have been reported.
Thus, understanding and precisely controlling ionic cores at a quantum level \cite{Muni} is of central importance for these applications.

In general, ionic-core transition is strongly perturbed by an outer electron due to non-negligible overlap of the wavefunction of Rydberg state with inner structure.
Excitation of an ionic core, in particular, leads rapid autoionization due to such coupling exhibiting a substantial broadening and a large amount of frequency shift in the excitation spectrum.
Thus, it hinders accurate control of ionic degrees of freedom in Rydberg atoms.
It is indeed found that such effects are reduced by bringing the Rydberg electron to a large orbit, namely high-$n$ and high-$\ell$ states, thereby isolating the ionic core from external disturbance.
Pioneering works on doubly-excited states of Rydberg atoms revealed scaling law of autoionization rate, $n^{-3}$ and $\ell^{-5}$, both theoretically and experimentally \cite{Cooke,Cooke1979,Cohen,Jones,Pruvost,Xu}.
Recent advances in controlling of Rydberg atoms and ionic cores \cite{Millen} also enabled very high-$n$ states \cite{Fields} and selective preparation of large-$\ell$ state deeply studied such scaling laws \cite{Lehec, Yoshida, Marin-Bujedo, Wehrli}.
Importantly, revisiting of circular Rydberg states realized an ideal platform for genuinely isolated-cores \cite{Holzl} demonstrating non-autoionizing system with almost lifetime limited spectrum \cite{Teixeira} and coherent coupling of outer-to-inner electrons \cite{Wirth}.

In this work, we developed a spectroscopic technique which enables significant suppression of autoionization, achieving more than 2 orders of magnitude reduction of linewidth of the ionic-core transition.
This is advantageous for gaining a high resolution in spectroscopy.
The key element of this method is population transfer to high-$\ell$ states with $\langle\ell\rangle>30$ via electric field control inspired by Stark switching and wavepacket technique \cite{Jones, Fields}, thus it converts the outer electron to a "spectator".
$\langle\ell\rangle$ denotes an expectation value of orbital quantum number of Rydberg electron. 
With this method, as a demonstration, we perform high-precision spectroscopy of an ionic-core transition, resolving isotope shifts and hyperfine splittings with previously unexplored accuracy.
Furthermore and equally importantly, we rigorously determine the frequency shift of the ionic-core transition.
To this end, we independently implemented a Paul trap to capture a single ion and we directly compare the spectrum of ionic cores to that of a single "bare" ion as is a highly accurate frequency reference.

We successfully determine the isotope shift and hyperfine splitting using the dipole transition, whose values are found to be in good agreement with those obtained for atomic ions \cite{Dubost, Barwood}, with an uncertainty of a few MHz.
Thus, the local properties of nuclear, namely, nuclear charge radii and magnetic dipole moment are similarly accessible for the case of ionic cores at comparable level of accuracy to atomic ions.
This fact simultaneously indicates that the ionic core is efficiently decoupled from the influence of the Rydberg electron.

\begin{figure}[t]
	\includegraphics[width=\columnwidth]{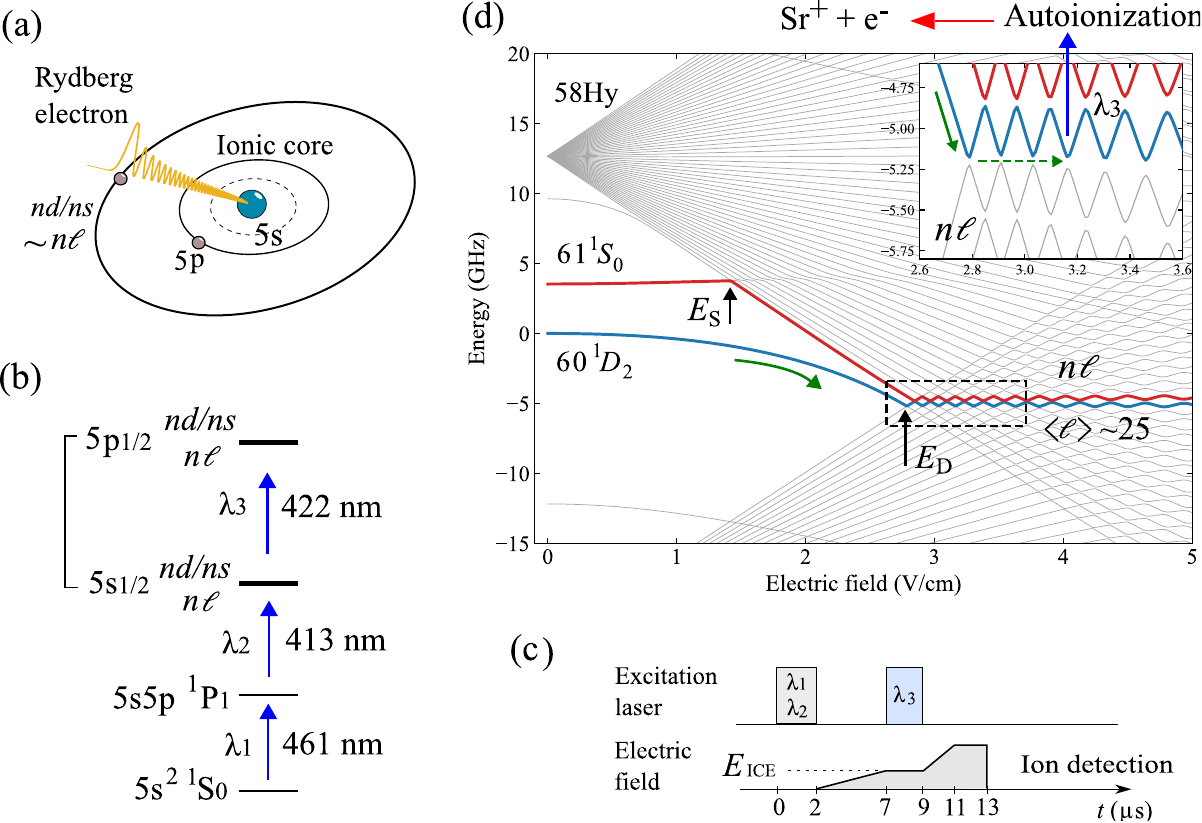}
	\caption{Ionic-core excitation and transfer of a Rydberg electron to high-$\ell$ states. (a) Schematic illustration of a Rydberg electron and the ionic core in a two-electron atom (Sr) is shown. (b) Relevant energy levels and excitation lasers are shown. The transition $5s_{1/2} n\ell \rightarrow 5p_{1/2} n\ell$ corresponds to a dipole transition of the ionic core. (c) Experimental sequence for the spectroscopy. $E_\mathrm{ICE}$ denotes the applied electric field during ionic-core excitation. (d) Calculated Stark structure in the vicinity of the $60^{1}D_{2}$ state ($m=0$) is plotted. The pathway for population transfer from $nD$ ($nS$) to $n\ell$ is indicated by a blue (red) line. $E_\mathrm{D}$ ($E_\mathrm{S}$) denotes the electric field at which the $nD$ ($nS$) state crosses the $n\ell$ manifold. The inset shows a magnified view of the intersection region (dashed box). The Rydberg electron is transferred to $n\ell$ state, and the subsequent core excitation is detected via autoionization.}
	\label{fig1}
\end{figure}

\begin{figure*}[t]
	\includegraphics[width=\textwidth]{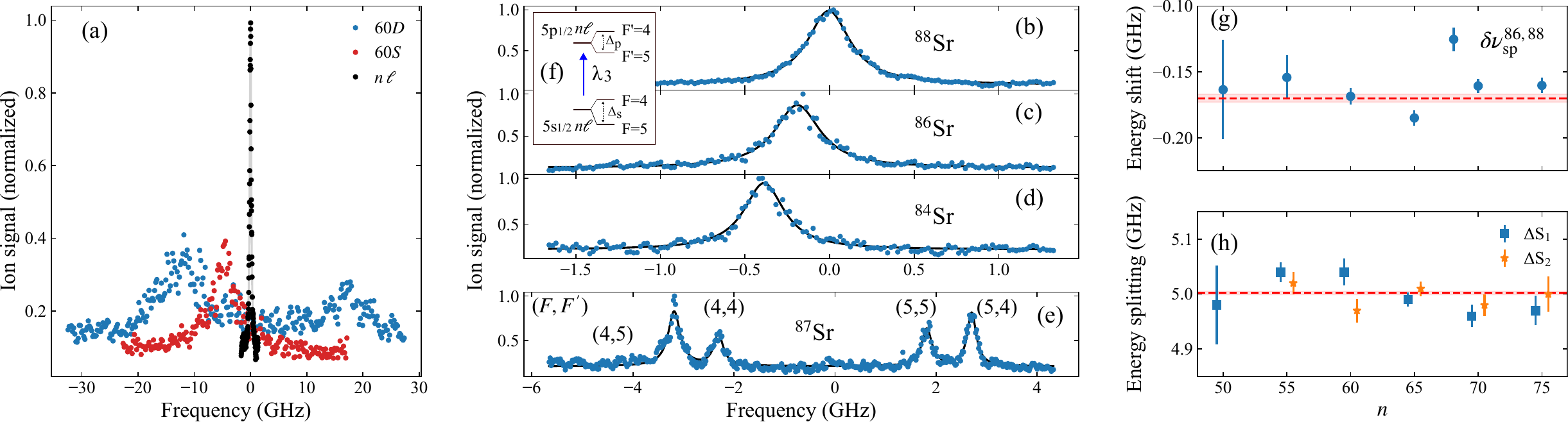}
	\caption{High-resolution spectroscopy of an ionic-core transition. (a) Ionic-core spectrum for Rydberg electron in 60$D$, 61$S$ and $n\ell$ states are shown. The linewidth for $n\ell$ state is $150\:\textrm{MHz}$. (b-e) Spectroscopy results for isotopes $^{88}$Sr, $^{86}$Sr, $^{84}$Sr and $^{87}$Sr resolving isotope shifts and hyperfine structure in ionic cores. Rydberg electron is prepared in $65\ell$. (f) Relevant hyperfine structure of the ionic core in $^{87}$Sr. (g,h) Isotope shift $\delta\nu^{86,88}_\textrm{sp}$ and hyperfine splitting $\Delta_\textrm{s}$ for various $n\ell$ are plotted. The error bar corresponds to statistical uncertainty (1$\sigma$). The dashed lines indicate previously reported values from atomic ions \cite{Dubost,Barwood}.}
	\label{fig2}
\end{figure*}


Our experiments consist of preparing ultracold samples of strontium isotopes and subsequent Rydberg and ionic-core excitation.
The details of the experimental setup are described in Supplemental Material.
To access an ionic-core transition, we first excite ground-state atoms to $n^{1}D_{2}$ or $n^{1}S_{0}$ Rydberg series via a two-photon process [see Fig.~\ref{fig1}(b)].
The experimental protocol is illustrated in Fig.~\ref{fig1}(c).
Subsequently, an electric-field pulse with a linear ramp profile (duration: $5\:\mu$s) is applied.
During the electric-field sweep, the population of Rydberg atoms initially in the $nD$ state is transferred (blue line and green arrow) toward a region where the $nD$ state connects to the upper and lower branches of neighboring $n\textrm{Hy}$ states.
After entering this region, the population cruises in a high-$\ell$ states [see the inset of Fig.~\ref{fig1}(d)].
At the final stage of population transfer, the Rydberg electron reaches $n\ell$ state with $\langle \ell \rangle \sim 25$.
Details for the calculation of $\langle \ell \rangle$ are given in Supplemental Material.
For convenience, we label high-$\ell$ state in the presence (absence) of electric field as $n\ell$ ($n\textrm{Hy}$).
Probability of diabatic transition to the neighboring levels at the avoided crossing is typically smaller than $3\times 10^{-3}$ calculated by using Landau-Zener formula \cite{Rubbmark}.
As soon as the electric field is set constant at $E_\mathrm{ICE}$, a third laser, $\lambda_{3}$, is applied for 2 $\mu$s, driving the ionic-core transition $5s_{1/2} n\ell \rightarrow 5p_{1/2} n\ell$.
Once the atoms are brought to the doubly excited state $5p_{1/2}n\ell$, autoionization produces strontium ions, which are subsequently guided and detected by a microchannel plate.
For spectroscopy, we scan the frequency of $\lambda_{3}$ and record the ion counts for each measurement.
This sequence is typically repeated 1000 times for each frequency setting to accumulate statistics.

We first present a summary of the main results of this work in Fig.~\ref{fig2}.
In this figure, we begin by showing linewidth squeezing of the ionic-core transition by electric-field control of Rydberg electron's orbit.
In Fig.~\ref{fig2}(a), we compare the ionic-core excitation spectra for Rydberg electrons prepared in the $60S$, $60D$, and $n\ell$ states.
For low-$\ell$ states, we observe a pronounced broadening effect caused by strong connection to autoionization states.
The spectra exhibit a wide signal spanning approximately $50\:\mathrm{GHz}$ associated with a significant frequency offset from the center.
Such a characteristic broadening and multiplet structure has also been reported in related studies \cite{Millen,Yoshida}.
In contrast, the spectral shape changes dramatically when the Rydberg electron is transferred to the $n\ell$ state, showing a single sharp peak with a linewidth of approximately $150\:\mathrm{MHz}$ (half width at half maximum).
This behavior reflects the efficient isolation of the ionic core from the outer electron when it is brought to a high-$\ell$ state, where the centrifugal barrier for high-$\ell$ shields penetration to the inner region.
This is essential for gaining a high resolution and precision in spectroscopy.
The dynamical evolution of the spectral shape and the associated linewidth narrowing are provided in detail later in this manuscript.

We now describe the high-resolution spectroscopy results of ionic cores using the technique explained above.
The representative data is shown in Fig.~\ref{fig2}(b-d) where we plot ionic-core spectrum for $^{88}\textrm{Sr}$, $^{86}\textrm{Sr}$, and $^{84}\textrm{Sr}$ with Rydberg electron prepared in $65\ell$.
We observe a distinct excitation peak for each isotope, and the relative shifts in the resonance frequencies, i.e., the isotope shifts, are clearly visible.
Fig.~\ref{fig2}(e) shows the result for $^{87}$Sr, an odd isotope with nuclear spin $I=9/2$.
Multiple peaks corresponding to the four possible transitions among the hyperfine levels are observed.
See Fig.~\ref{fig2}(f) for the relevant energy levels.
All the spectrum are well fitted by a Lorentzian function, indicating that it is governed by residual autoionization, an intrinsic property of doubly excited states.

In order to test the stability of the ionic core transitions, we further proceed the spectroscopy by changing $n$ of the Rydberg states.
Observed isotope shift and the hyperfine splitting for $50\ell-75\ell$ is plotted in Fig.~$\:$\ref{fig2} (g,h).
Typical $E_\textrm{ICE}$ used in this measurement is $\sim\:1.2\:\times\:E_\textrm{D}$.
The extracted isotope shift $\delta\nu^{86,88}_\textrm{sp}\equiv\nu^{86}_\textrm{sp}-\nu^{88}_\textrm{sp}$ is determined as $-172(16)\:\textrm{MHz}$ (mean value with standard error) for $^{86}\textrm{Sr}$ and $^{88}\textrm{Sr}$.
Here, $\nu^{86(88)}_\textrm{sp}$ denotes resonance frequency of the ionic core transition in $^{86(88)}\textrm{Sr}$ atom.
This shows a reasonable agreement with a previously reported value $-170(3)\:\textrm{MHz}$ obtained from measurement with an ion crystal \cite{Dubost} [dashed line in Fig.~\ref{fig2}(g)].
Similarly, the shift for $^{84}\textrm{Sr}$ and $^{88}\textrm{Sr}$ is derived as $\delta\nu^{84,88}_\textrm{sp}$=$-387(6)\:\textrm{MHz}$ comparable to the value from the same report \cite{Dubost}.
The summary of the isotope shift as well as hyperfine splitting determined in this work is shown in Table~$\:$\ref{tab1}.
The hyperfine splitting of $^{87}\textrm{Sr}$ ionic core is extracted by measuring the frequency spacing of multiple peaks in Fig.~$\:$\ref{fig2}(e).
The result is plotted in Fig.~$\:$\ref{fig2}(h).
$\Delta_{s1}$ ($\Delta_{s2}$) represents the frequency difference of $(F,F')=(4,5)$ and $(5,5)$ [$(4,4)$ and $(5,4)$], which is equivalent to the hyperfine splitting in $5s_{1/2}$ state.
$\Delta_{s1}$ and $\Delta_{s2}$ are horizontally shifted on purpose for visibility.
The determined hyperfine splitting in the ground state, $\Delta_\textrm{s}\:=\:4996(26)\:\textrm{MHz}$, shows a reasonable agreement with the value obtained from a work using atomic ions \cite{Barwood}.
The hyperfine constant is derived as $A=-999(5)$ which compares with the value from \cite{Sunaoshi}.
The hyperfine splitting of the upper state $5p_{1/2}$ is similarly obtained.
The corresponding data is given in Supplemental Material.
In Fig.~$\:$\ref{fig2}(g,h), we indeed observe a larger uncertainty and scattering of the data point towards the lower $n$.
This originates from line broadening in the spectrum governed by the overlap of wave functions in the radial direction with the inner structure.
We investigate this feature in depth later in the next section.

\begin{table}[b]
\caption{\label{tab1}%
Summary of isotope shifts and hyperfine splitting of ionic-core transitions in strontium isotopes.
}
\begin{ruledtabular}
\begin{tabular}{lcr}
\textrm{Shift/Splitting}&
\textrm{Ionic core (this work) (MHz)}&
\textrm{Atomic ion (MHz)}\\
\colrule
$\delta\nu^{86,88}_\textrm{sp}$ & -172(16)  & -170(3) \cite{Dubost}\\
$\delta\nu^{84,88}_\textrm{sp}$ & -387(6) & -378(4) \cite{Dubost}\\
$\Delta_\textrm{s} (^{87}\textrm{Sr})$ & 4996(26) &  5002.30(5) \cite{Barwood}\\
$\Delta_\textrm{p} (^{87}\textrm{Sr})$ & 888(2) &  892 (calc.) \cite{Yu, Martensson-Pendrill}\\
\end{tabular}
\end{ruledtabular}
\end{table}

\begin{figure*}[t]
	\includegraphics[width=\textwidth]{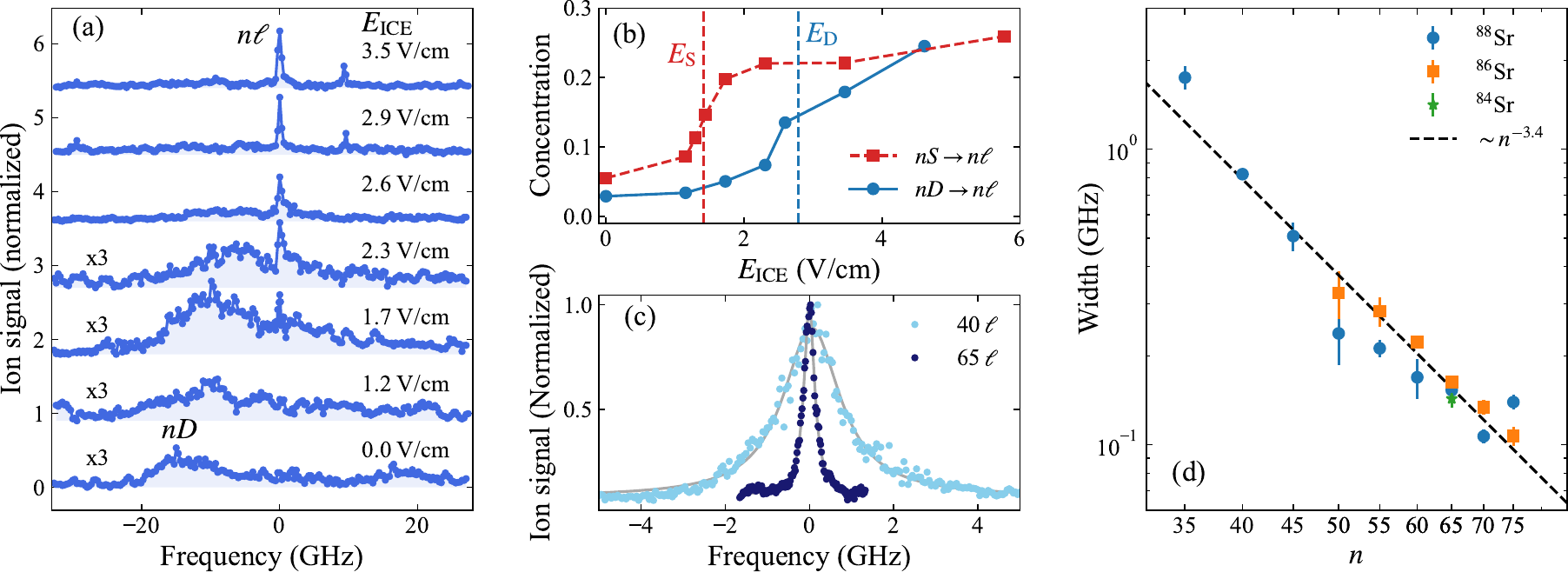}
	\caption{Evolution of the ionic-core excitation spectrum under transfer from $nD$ to $n\ell$ state and $n$-scaling of the linewidth. (a) Wide-range scan across the $5s_{1/2} \rightarrow 5p_{1/2}$ resonance for various $E_\mathrm{ICE}$. The signal amplitude is magnified by a factor of 3 for $E_\mathrm{ICE} < 2.5\:\mathrm{V/cm}$ for visibility. The broad component of the ion signal in the $nD$ state is progressively concentrated toward the central peak as $E_\mathrm{ICE}$ is increased. The origin of a small peak at $9.5\:\mathrm{GHz}$ has not yet been identified. (b) Concentration in the central component as a function of $E_\mathrm{ICE}$. $E_\textrm{S}$ and $E_\textrm{D}$ corresponds to the crossing point of $61S$ and $60D$ with $n\:\ell$ branches [see Fig.~$\:$\ref{fig1}(d)]. (c) Comparison of spectra for the $40\ell$ and $65\ell$ states with Lorentzian fit ($^{88}$Sr).(d) Extracted linewidth of $5s_{1/2}n\ell - 5p_{1/2}n\ell$ transition is plotted as a function of $n$. Results for $^{88}$Sr, $^{86}$Sr, and $^{84}$Sr are included. The dashed line represents a fit with a power-law scaling $\sim n^{\alpha}$, with $\alpha = -3.4$.}
	\label{fig3}
\end{figure*}

As explained, the spectral shape of the ionic-core transition is strongly influenced by the orbital of the Rydberg electron.
In the following, we examine the dynamical deformation of the spectral shape during population transfer from low-$\ell$ to $n\ell$ states.
The results are shown in Fig.~\ref{fig3}.
We perform spectroscopy for various electric field $E_\textrm{ICE}$ at core excitation [See Fig.~\ref{fig3}(a)].
Starting from a pure $60^{1}D_{2}$ state in $^{88}$Sr ($E_\textrm{ICE}=0\:\textrm{V/cm}$), a gradual shift of the broad spectrum toward the resonance center is observed as $E_{\mathrm{ICE}}$ is increased.
When $E_{\mathrm{ICE}}$ reaches $1.7\:\mathrm{V/cm}$, a sharp peak emerges at the center frequency region.
The height of this peak rapidly grows whereas the signal in the broad component is redistributed to the sharp central peak ($n\ell$) by increasing $E_\textrm{ICE}$.
At a final stage, it dominates at sufficiently large values of $E_{\mathrm{ICE}}$.
This behavior is an indication of the isolation of the ionic core via transfer from $nD$ to $n\ell$ by mitigating autoionization.
Similar dynamics are indeed also confirmed for the case of the initially prepared state in $60^{1}S_{0}$.
The result and analysis for this state is provided in Supplemental Material.
We further evaluate the population transfer process by analyzing the concentration of the ion yield into the central peak.
The result is plotted in Fig.~\ref{fig3}(b) where we find a general trend of growth of concentration as $E_{\mathrm{ICE}}$ is increased for both cases of $nD \rightarrow n\ell$ and $nS \rightarrow n\ell$.
The concentration is here defined as $A_{\mathrm{c}}/A_{\mathrm{tot}}$, where $A_{\mathrm{c}}$ and $A_{\mathrm{tot}}$ represent the area of the central component and the total signal (shaded area over the full frequency range), respectively \cite{Concentration}.
Notably, we observed that the onset of the rapid signal jump occurs for $nS \rightarrow n\ell$ at lower $E_{\textrm{ICE}}$, compared to $nD \rightarrow n\ell$ case.
This difference is due to the fact that the crossing point of $61S$ with the high-$\ell$ state, namely $E_{\mathrm{S}} = 1.42\:\mathrm{V/cm}$, is located at a lower electric field than $E_{\mathrm{D}} = 2.78\:\mathrm{V/cm}$ for the $60D$ state [See Fig.~\ref{fig1}(d)].
As a result, the coupling to the $n\ell$ manifold occurs earlier (later) for $60S$ ($60D$) in electric-field amplitude.
We suggest the more gentle slope for signal rise in $60D$ state is attributed to slight variation of $E_\textrm{D}$ for different $m$ ($m=0,1,2$).
The electric-field required for signal rise for both $nS$ and $nD$ reasonably matches with $E_{\mathrm{S}}$ and $E_{\mathrm{D}}$ [dashed lines with red and blue in Fig.~\ref{fig3}(d)].
We further confirm that both cases end up to a similarly narrow peak at sufficiently large values of $E_{\mathrm{ICE}}$.
This result is consistent with the transient pathways of the state transfer among the Stark structure, as introduced earlier in this manuscript.

We now focus on the central peak, in particular, addressing the residual broadening effect.
In Fig.~\ref{fig3}(c), we plot representative data for ionic-core spectrum for the $40\ell$ and $65\ell$ states to highlight the difference in linewidth.
To understand the underlying mechanism of the broadening, we further perform similar measurements over a wide range of $n$.
The extracted linewidth for various $n$ is summarized in Fig.~\ref{fig3}(d).
Measured data on different isotopes are also included in the figure.
The data reveals a power-law scaling of the linewidth with $n$, given by $n^{-3.4}$ from a fitting.
This scaling is mainly governed by the overlap of wavefunction in the radial direction.
Similar dependence of autoionization rate on $n$ is also found for a fixed low-$\ell$ case as $n^{-3}$ \cite{Cooke, Fields}.
A possible explanation for the slightly faster decay observed in the current work comes from the differences in $\langle\ell\rangle$ for different $n$.
Specifically, higher-$n$ states contain a larger fraction of high-$\ell$ components (c.f. autoionization rate has a stronger dependence on $\ell$ as $\ell^{-5}$ as discussed in \cite{Marin-Bujedo, Yoshida, Jones}).

\begin{figure}[h]
	\includegraphics[width=\columnwidth]{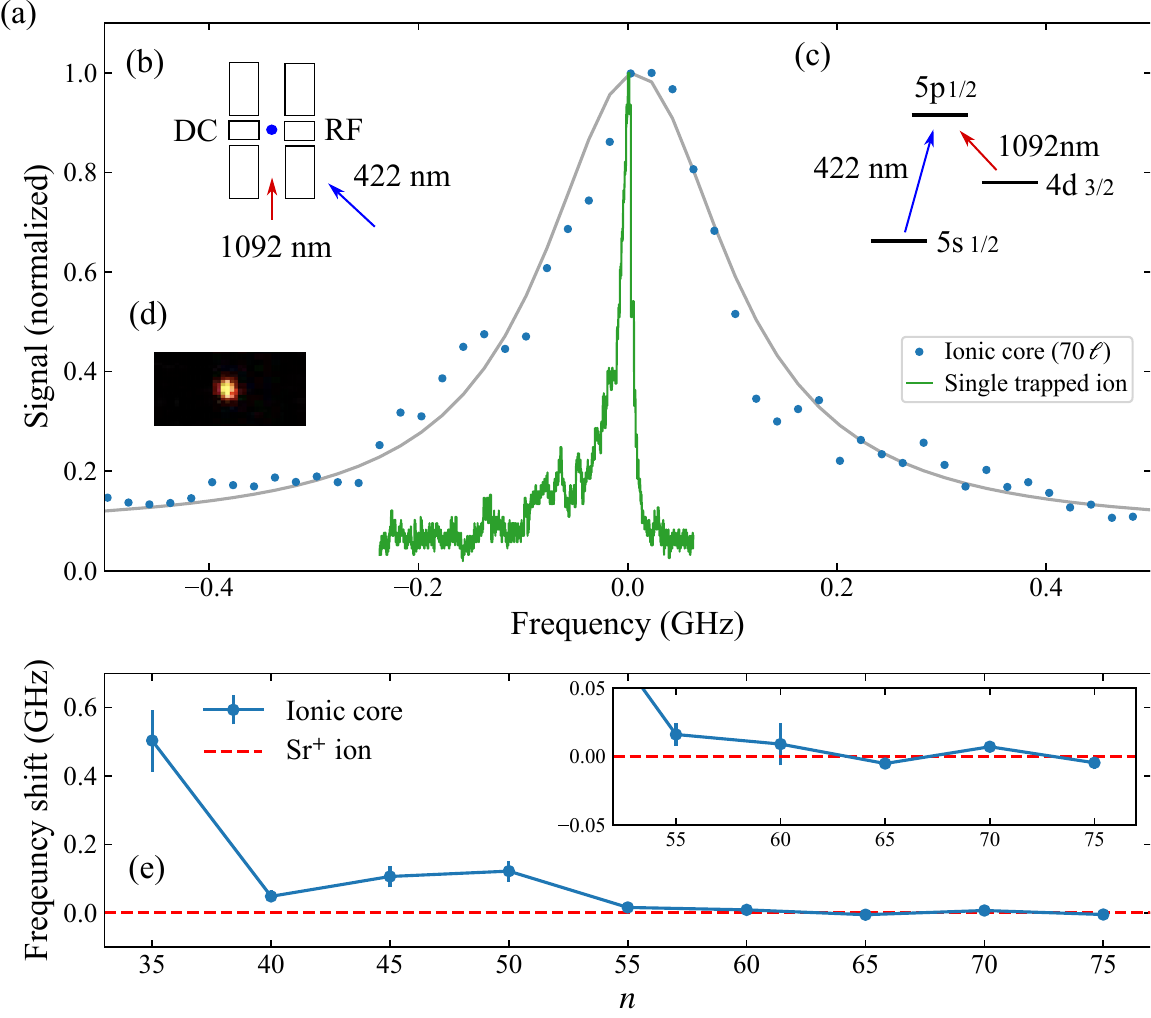}
	\caption{Direct comparison of ionic-core spectrum with a single ion's signal. (a) Ionic-core excitation ($70\ell$) is plotted together with the fluorescence signal from a single $^{88}$Sr$^{+}$ ion. (b) The schematic of the linear Paul trap is shown. (c) Relevant energy diagram of Sr$^{+}$ ion is illustrated. (d) An image of a single ion. (e) Frequency shift vs. $n$. The inset represents a zoom-in picture of data for $n=55 - 75$ region.}
	\label{fig4}
\end{figure}

So far, we have mainly discussed the linewidth of the ionic-core transitions; however, the frequency shift is also a critical quantity that determines the accuracy of spectroscopic measurements.
In order to unambiguously specify the frequency shift, we directly compare the spectrum with the one from a single trapped ion, serving as an ultimate frequency reference.
To this end, we implemented a Paul trap to confine a single Sr$^{+}$ ion as an independent spectroscopy platform.
The apparatus is schematically illustrated in Fig.~\ref{fig4}(b).
In brief, the trap is a conventional linear Paul trap consisting of DC and RF electrodes.
A detailed description of the ion trap is provided in Supplemental Material.
We apply a $33\:\mathrm{MHz}$ oscillating field with an amplitude of $\sim 1\:\mathrm{kV}$ to the RF electrodes.
For axial confinement, two endcaps are used.
The resulting trapping frequencies are $(\omega_\textrm{r}, \omega_\textrm{ax}) = 2\pi \times (1890, 590)\:\mathrm{kHz}$, thereby eliminating Doppler shifts and residual broadening in the resonance spectrum.
Here, $\omega_{\textrm{r/ax}}$ denotes the trapping frequency in the radial and axial direction, respectively.
Fig.~\ref{fig4}(d) shows an image of a single $^{88}$Sr$^{+}$ ion taken with an electron-multiplying charge-coupled device camera.
To perform spectroscopy of a trapped ion, we use the same laser source as in the Rydberg-atom setup ($\lambda_{3}$ laser).
The measurement result of the $5s_{1/2} \rightarrow 5p_{1/2}$ transition of a single $^{88}$Sr$^{+}$ ion is shown in Fig.~\ref{fig4}(a).
The data (green solid line) represent fluorescence from the ion detected by a photomultiplier tube while scanning the $\lambda_{3}$ laser, exhibiting a lifetime-limited linewidth ($\Gamma=2\pi\times20.3\:\mathrm{MHz}$).
The asymmetric line shape of the ion spectrum, i.e., the abrupt drop of the signal on the blue-detuned side, originates from heating effects.
An additional laser at $1092\:\mathrm{nm}$ is used to repump population from the metastable $4d_{3/2}$ state.
See Fig.~\ref{fig4}(c) for the relevant energy levels.
We plot the result from the ionic-core ($70\ell$) as blue dots for comparison.
This comparison reveals the frequency shift, $\nu_\textrm{core}-\nu_\textrm{ion}=+7.0(3.1)\:\mathrm{MHz}$.
$\nu_\textrm{core}$ and $\nu_\textrm{ion}$ denotes resonant frequency of a core ion and a trapped ion, respectively.
We perform similar comparison for the spectrum with different $n\ell$ states.
The result is shown in Fig.~\ref{fig4}(e), where the frequency shift is plotted as a function of $n$.
We observe that the shift is well stabilized to be a small value well within $30\:\mathrm{MHz}$ for $n>50$ [see inset in Fig.~\ref{fig4}(e)], as is suggesting that the quantum defect of the doubly-excited state is rather small \cite{Jones}.
The frequency shift increases rapidly to approximately $+500\:\mathrm{MHz}$ toward lower $n$.
Validation of the frequency shift in this manner ensures the reliability of the experimental results throughout this study, for instance, determination of the isotope shift, especially for a domain $n>50$ where the frequency shift has minimal impact.

The final state as a consequence of population transfer in this work is a highly mixed state within the high-$\ell$ manifold, therefore, core-ion excitation is in principle expected to involve multichannel transitions.
Thus, it can lead spectral broadening, particularly in the presence of an electric field.
This is considered as one of the factors limiting the minimum achievable linewidth.
Theoretical prediction for autoionization rates for high-$\ell$ states by \cite{Marin-Bujedo}, where rates for $\ell=0-45$ were calculated, yields values orders of magnitude smaller than those observed in our experiment.
A possible source for this discrepancy may arise from the multichannel effect discussed above.
We examined the impact of the electric field on the linewidth of the ionic-core transition and indeed observed a slight variation of the linewidth for different $E_{\mathrm{ICE}}$.
In addition, fluorescence decay from $5p_{1/2}$ state becomes increasingly relevant under conditions where the autoionization is strongly suppressed.
A more detailed experimental and theoretical investigation on the multichannel effect and interplay with florescence decay as well as relaxation to neighboring Rydberg states via black-body radiation is required to elucidate these aspects.
Nevertheless, the method developed here proves to be useful for substantial suppression of autoionization even for a relatively low magnetic quantum numbers $m$.
For ultimate precision, realization of a non-autoionizing ionic core via circular states with the maximum possible $m$ quantum number \cite{Teixeira, Holzl} or preparing extremely high-$n$ states \cite{Fields} provide a more suitable platform.

In conclusion, we reported precise test of the dipole transition in ionic cores by examining the isotope shifts and hyperfine splitting as a demonstration of the performance.
The resulting linewidth obtained in this work reaches a factor of 5 to a natural linewidth of an atomic ion.
The uncertainty is at a level of a few to ten MHz in the current study.
The isotope shifts and hyperfine splittings determined in this work show no discernible deviation from the results obtained for atomic ions, demonstrating the power of the spectroscopy technique combined with direct comparison to a single ion signal developed here.
As a result, we confirmed that the core ions are effectively isolated and become more alike a bare ion.
Although the achievable linewidth in this work is limited to $10-100\:\textrm{MHz}$ regime due to the intrinsic character of strong dipole transitions, the method reported here is readily extended to other types of transitions.
More interesting application is a quadrupole transition.
Much narrower nature of dipole-forbidden transitions offers unprecedented level of accuracy in spectroscopy, and ultimately, quantum control of ionic cores and sensitive probing of electron-core interaction is expected to be realized.

\begin{acknowledgments}
We acknowledge insightful discussion with Masahiro Takeoka. 
This work is supported by Japan Science and Technology Agency Moonshot R \& D Grant No. JPMJMS2063, No. JPMJMS256G and PRESTO Grant No. JPMJPR2459.
\end{acknowledgments}

\bibliographystyle{apsrev4-2}
\bibliography{refs}

\section*{Supplemental Material}

\subsection{Experimental setup}

\begin{figure}[h]
	\includegraphics[width=\columnwidth]{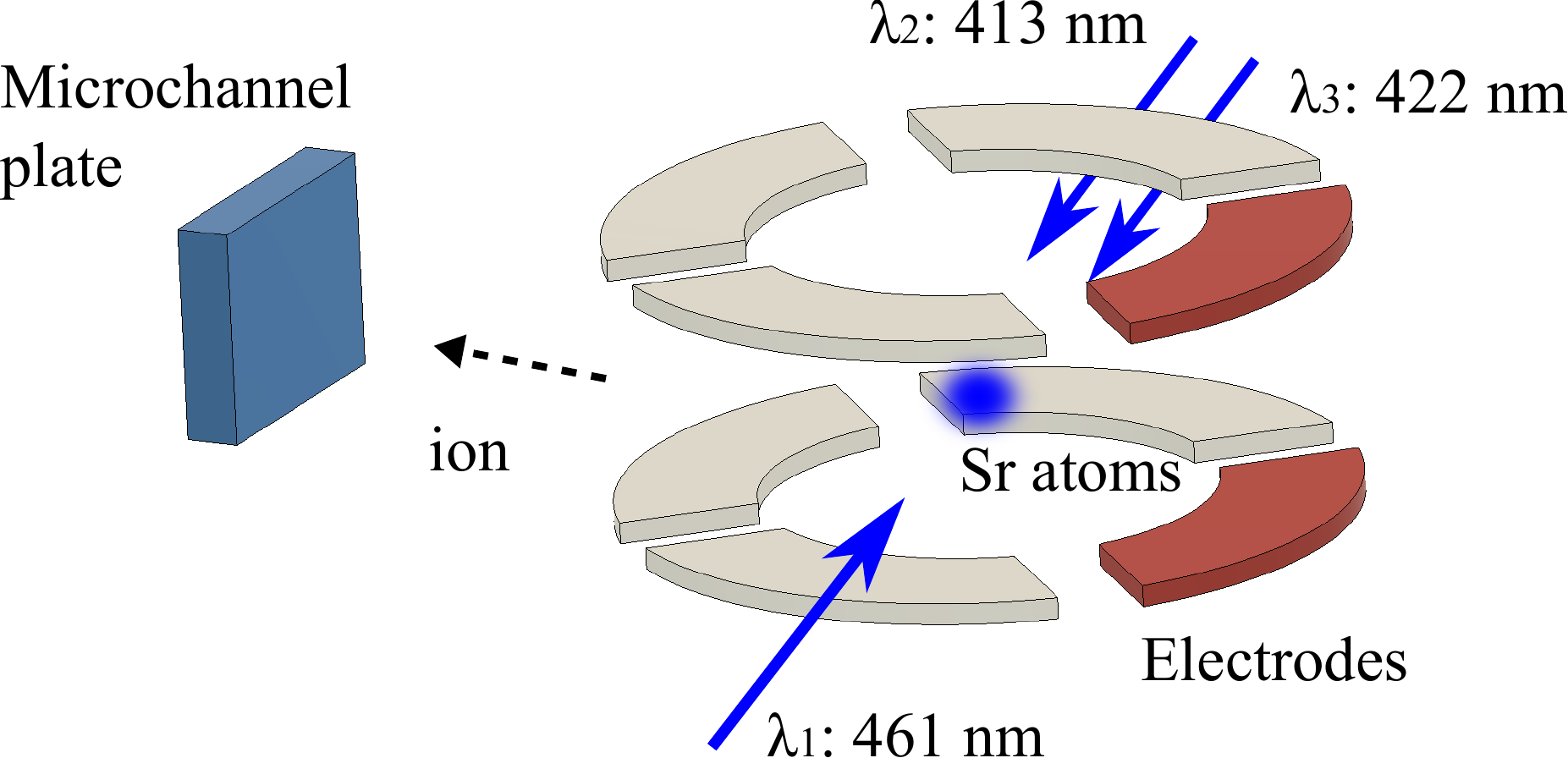}
	\caption{Experimental setup. Strontium atoms are trapped in a magneto-optical trap. Three lasers, $\lambda_{1,2,3}$, are used for excitation of cold atomic ensemble to Rydberg states and subsequent excitation of ionic-core transitions. Segmented-shaped plates are electrodes for controlling electric field in experiment. A blue square indicates microchannel plate for ion detection.}
	\label{figS1}
\end{figure}

Strontium isotopes, $^{88}$Sr, $^{87}$Sr, $^{86}$Sr, and $^{84}$Sr, are laser cooled and selectively loaded in a magneto-optical trap.
The experimental setup is shown in Fig.~\ref{figS1}.
For cooling, the $^{1}$S$_{0} \rightarrow {}^{1}$P$_{1}$ transition at $461\:\mathrm{nm}$ is used.
Lasers at $707\:\mathrm{nm}$ and $697\:\mathrm{nm}$ are also applied for repumping from metastable triplet states.
The typical number of atomic samples is $10^{6}$-$10^{8}$, depending on the isotope.
Two laser beams, $\lambda_{1}$ and $\lambda_{2}$, at $461\:\mathrm{nm}$ and $413\:\mathrm{nm}$, are used to excite ground-state atoms to Rydberg states via the $^{1}P_{1}$ intermediate state.
The frequency of $\lambda_{1}$ is detuned by $-450\:\mathrm{MHz}$ from the $^{1}S_{0} \rightarrow {}^{1}P_{1}$ resonance.
The second laser at $413\:\mathrm{nm}$ ($\lambda_{2}$), counter-propagating with respect to $\lambda_{1}$, excites to the $n^{1}D_{2}$ or $n^{1}S_{0}$ Rydberg series.
$\lambda_{1,2,3}$ laser sources are all continuous wave lasers and are pulsed  using acousto-optic modulators.
The third laser $\lambda_{3}$ at $422\:\mathrm{nm}$ drives the ionic-core transition $5s_{1/2} n\ell \rightarrow 5p_{1/2} n\ell$.
Excitation of core excitation is detected via autoionization and the produced ions are guided to a microchannel plate located approximately $10\:\mathrm{cm}$ from the atomic sample, where ion signals are recorded.
Segmented electrodes are used to control the electric field at the position of the atomic ensemble.
In the measurement sequence, electric-field pulse for population transfer and ion extraction is applied to designated electrodes (red colored).
The measurement protocol is shown in Fig.~\ref{fig1}(c) in the main text.

\subsection{Starkmap in the vicinity of $60^{1}D_{2}$}

\begin{figure}[h]
	\includegraphics[width=\columnwidth]{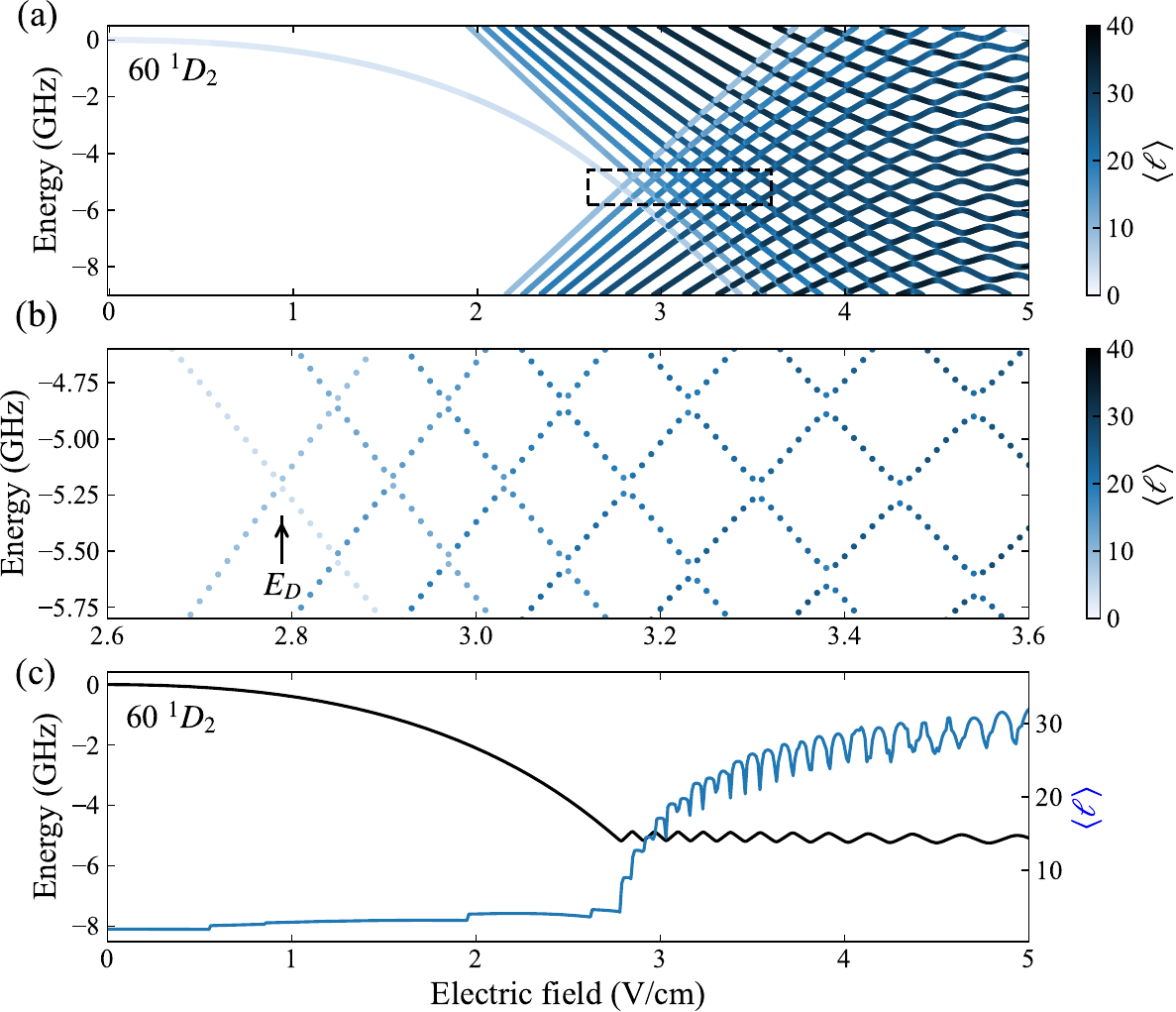}
	\caption{Calculated Starkmap in the vicinity of $60^{1}D_{2}$ ($m=0$). (a) Energies for each levels are plotted with weighted color according to $\langle\ell\rangle$. (b) Zoomed picture of the dashed box. (c) $\langle\ell\rangle$ of the state with initial composition of pure $60^{1}D_{2}$ state (blue line). The energy of the corresponding state is also shown.}
	\label{figS2}
\end{figure}

We calculate the energy levels of Rydberg state in the presence of electric field.
The calculated energy structure, Starkmap, is shown in Fig.$\ref{figS2}$.
For calculation, we use Alkali Rydberg Calculator \cite{Sibalic}, a python package for calculating Rydberg atoms, with a slight customization.
The eigen energies in the vicinity of $60^{1}D_{2}$ and a high-$\ell$ region are plotted ($m$=0).
Each point is highlighted by a color according to calculated expectation values of $\ell$, where $\langle\ell\rangle$=
$\sum_\ell
\ell
 \sum_{n,j,m}
\left|
\langle n \ell jm \mid \Psi \rangle
\right|^2.$

Fig.$\ref{figS2}$(b) is a zoomed-in picture of intersectional region of $60^{1}D_{2}$ and $n\ell$ states [the dashed box in Fig.$\ref{figS2}$(a)].
In Fig.$\ref{figS2}$(c), we plot $\langle\ell\rangle$ of the state starting from pure $60^{1}D_{2}$ at vanished field as well as its eigen energy.
When electric field is above the crossing point $E_\textrm{D}$, $\langle\ell\rangle$ rapidly increased over $\ell\:>\:25$ for $n=60$.
The oscillatory behavior is due to avoided crossings to neighboring levels.

\subsection{Hyperfine splitting of $5p_{1/2}$ in an ionic core}

Here, we plot the measurement result of hyperfine splitting $F'=4 - F'=5$ of $5p_{1/2}n\ell$ state in $^{87}$Sr$^{+}$.
In order to derive the splitting, we analyze the spectrum shown in Fig.\ref{fig2}(e).
We extract the energy spacings, $\Delta_\textrm{p1}$ and $\Delta_\textrm{p2}$, by subtracting the resonance frequency of $(4,4)-(4,5)$ and $(5,4)-(5,5)$ transitions among the four spectrum labeled with $(F,F')$.
See Fig.\ref{fig2}(f) for the relevant energy levels.
In the plot, we deliberately shift $\Delta_\textrm{p1}$ and $\Delta_\textrm{p2}$ horizontally for visibility.
The mean value for the data from $n=45-75$ is $888\:(2)\:\textrm{MHz}$.
The corresponding hyperfine structure constant $A=-177.6(4) \textrm{MHz}$ agrees with the calculated value $-178 \textrm{MHz}$ \cite{Yu, Martensson-Pendrill} which is plotted as a dashed line for comparison.

\begin{figure}[h]
	\includegraphics[width=\columnwidth]{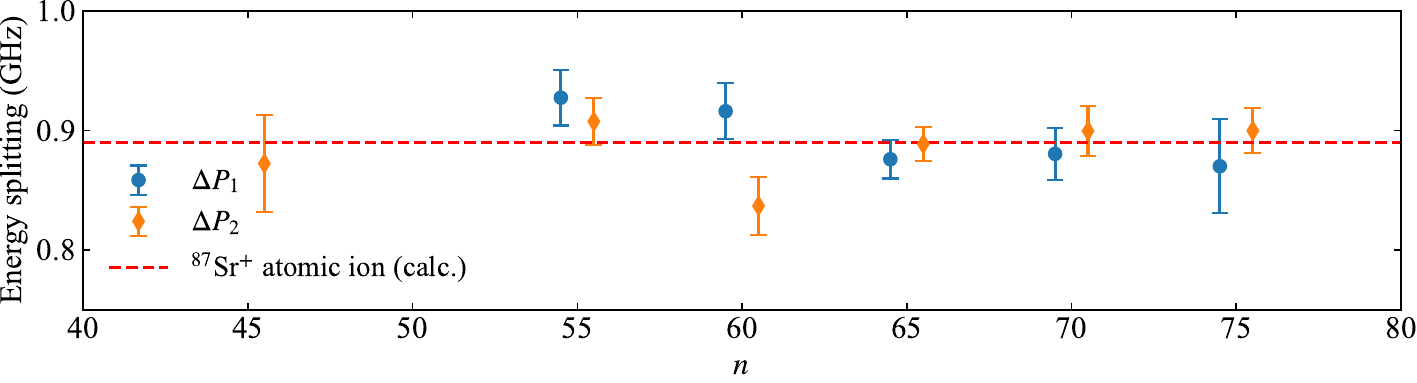}
	\caption{Measured hyperfine splitting of $5p_{1/2}n\ell$ in ionic core of $^{87}$Sr. Extracted energy splitting of $F'=4 - F'=5$ are plotted as a function of $n$. Dashed line is a calculated value of $^{87}$Sr$^{+}$ \cite{Yu,Martensson-Pendrill}}
	\label{figS3}
\end{figure}

\subsection{Population transfer from $nS$ to $n\ell$ state}

\begin{figure}[h]
	\includegraphics[width=\columnwidth]{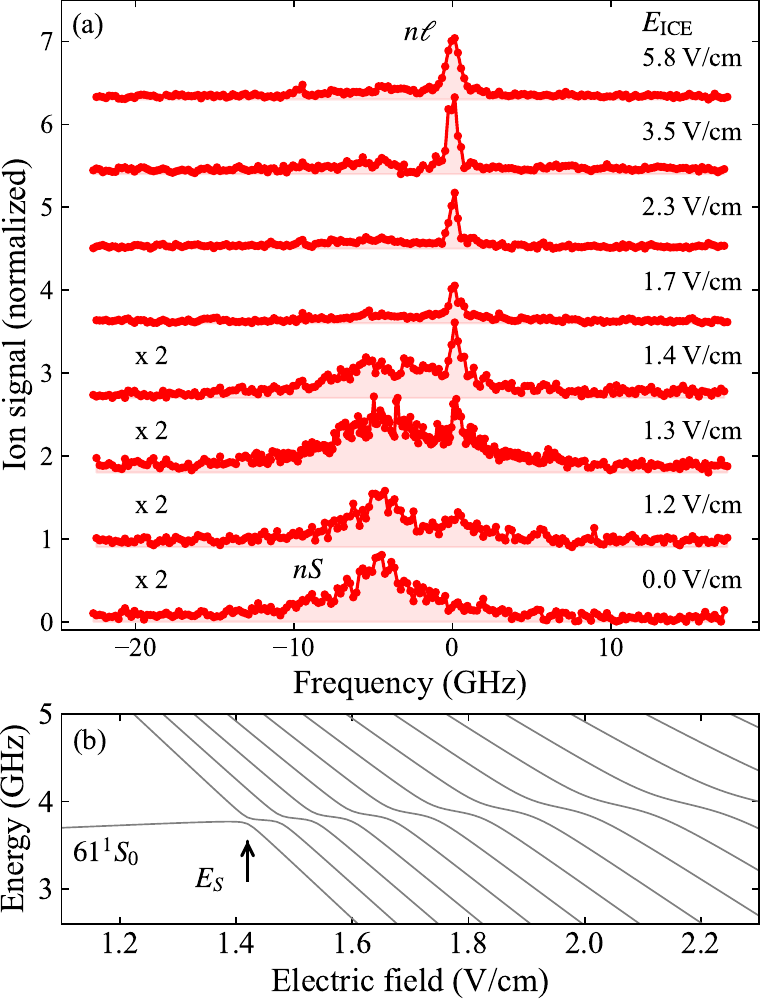}
	\caption{Spectral shapes of ionic-core transition under $60^{1}S_{0}\rightarrow n\ell$ transfer. (a) Spectrum of ionic-core transition for $E_\textrm{ICE}$= 0 to $5.8\:\textrm{V/cm}$  are shown. Ion signals for $E_\textrm{ICE}<1.5\:\textrm{V}$ is doubled for visibility. (b) Calculated Starkmap in the vicinity of $61^{1}S_{0}$ state is shown.}
	\label{figS4}
\end{figure}

We tested the impact of population transfer of Rydberg electron on the spectral shape of ionic-core transition, staring from $61^{1}S_{0}$ state and transferring to $n\ell$ as well.
Thus, we can confirm the consistency with the result for $60^{1}D_{2}$ state to $n\ell$ described in the main text.
The measured data is summarized in Fig.\ref{figS4}.
A similar experimental protocol as is introduced in Fig.\ref{fig3} (a) is used except the Rydberg electron is initially prepared in $61^{1}S_{0}$ state.
For $E_\textrm{ICE}=0$, a broad resonance with $\sim5\:\textrm{GHz}$ width is seen as a single component.
Gradual increase of $E_\textrm{ICE}$ reduces the population in the broad component whereas a growth of a central component is visible.
When $E_\textrm{ICE}$ goes above $E_\textrm{S}=1.42\:\textrm{V/cm}$ (see a figure below), ion signals concentrate on the narrow component and further increase of the electric field saturates the signal.
This behavior is plotted in Fig.\ref{fig3}(b) in the main text.
The signal concentration to the central peak occurs in lower $E_\textrm{ICE}$ compared with the case starting from $60^{1}D_{2}$, since the crossing for $61^{1}S_{0}$ with $n\ell$ manifolds comes at lower electric field at $E_\textrm{S}$ [See Fig.~\ref{fig1}(d) in the main text].
Fig.\ref{figS4}(b) is calculated Starkmap around $61^{1}S_{0}$ state.

\subsection{Paul trap}

\begin{figure}[h]
	\includegraphics[width=\columnwidth]{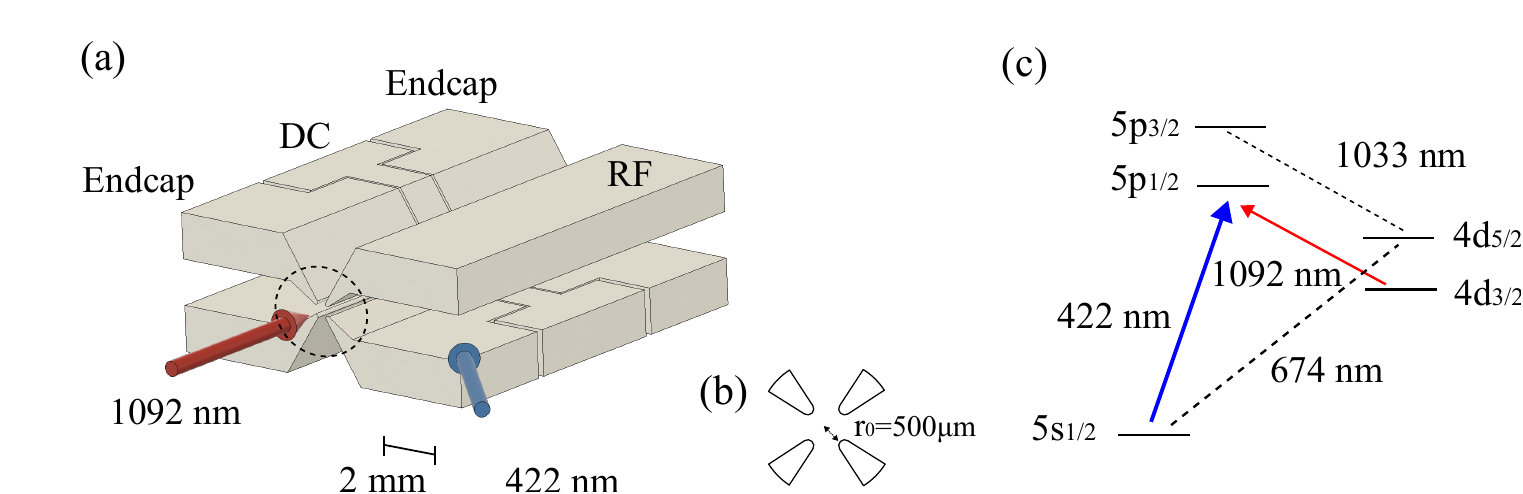}
	\caption{Paul trap. (a) The design of the Paul trap with segmented electrodes is shown. (b) The zoomed picture of dashed circle showing a radial profile of the trap. The ion-electrode distance is $r_{0}=500\: \mu \textrm{m}$. (c) Energy level diagram of $^{88}$Sr$^{+}$ ion is shown.}
	\label{figS5}
\end{figure}

The design of the implemented Paul trap, a conventional linear trap, is schematically shown in Fig.~\ref{figS5}(a).
RF and static voltages are applied to segmented electrodes for three-dimensional confinement of a Sr$^{+}$ ion.
The frequency of RF signal is $33\:\textrm{MHz}$ with an amplitude of maximally $1000\:\textrm{V}$.
Static voltage, typically $310\:\textrm{V}$, is applied to endcap electrodes for axial confinement.
The resulting trapping frequency for a single Sr$^{+}$ is $(\omega_{r},\omega_{ax})=2\:\pi\times(1890,590)\:\textrm{kHz}$ measured by sideband-resolved spectroscopy using $5s_{1/2}-4d_{5/2}$ quadrupole transition at 674 nm.
See Fig.\ref{figS5}(c) for the relevant energy level diagram.
To load a single ion, we heat an effusive oven for a few minutes.
Simultaneously, we shine lasers at $422\:\textrm{nm}$ and $1092\:\textrm{nm}$ for laser cooling together with $461\:\textrm{nm}$ and $405\:\textrm{nm}$ lasers for photoioniaztion.
$422\:\textrm{nm}$ laser is shared with the spectroscopy experiment of ionic cores in Rydberg atoms.
Fluorescence signal during doppler cooling is collected through an objective lens and detected by a photomultiplier tube as well as an imaging camera above (not shown).
Typical photon count from a single ion is $2\times10^{5}\textrm{count/s}$.
The fluorescence data shown in Fig.\ref{fig4}(a) is recorded while scanning the frequency of $422\:\textrm{nm}$.
$1092\:\textrm{nm}$ laser is kept on during the measurement for repumping from metastable $4d_{3/2}$ state.

\end{document}